\documentclass[12pt]{article}

\usepackage{graphics}
\usepackage{amsmath}
\def\ni{\noindent}
\def\ph{{\phantom{...}}}
\def\={\phantom{..} = \phantom{..}}

\def\siml{\phantom{..} \sim \phantom{..}}
\def\+{\phantom{..} + \phantom{..}}
\def\>{\phantom{..} > \phantom{..}}
\def\<{\phantom{..} < \phantom{..}}
\def\-{\phantom{..} - \phantom{..}}

\def\bq{\begin{quote}}
\def\eq{\end{quote}}
\def\be{\begin{equation}}
\def\ee{\end{equation}}
\def\bar{\begin{eqnarray}}
\def\ear{\end{eqnarray}}
\def\no{\nonumber}
\def\pa{\partial}

\def\MA{Mathematical Appendix}

\def\ea{{\em et. al}}

\def\Sch{Schr{\"o}dinger}
\def\Schism{Schr{\"o}dingerism}
\def\Schist{Schr{\"o}dingerist}
\def\Schists{Schr{\"o}dingerists}

\def\Schs{Schr{\"o}dinger's}
\def\Schseqn{Schr{\"o}dinger's equation}

\def\md{macroscopic dispersion}
\def\vN{von Neumann}

\def\Ham{Hamiltonian}
\def\wf{wavefunction}

\def\MP{Measurement Problem}
\def\qm{quantum mechanics}
\def\NLQM{Non-Linear Quantum Mechanics}
\def\MQP{macroscopic quantum phenomena}

\def\plusp#1{$10^{#1}$}
\def\pluspp#1#2{$10^{#1#2}$}

\def\negpp#1#2{$10^{-#1#2}$}

\def\cHqm{{\bf{H}_{QM}}}
\def\cHnl{{\bf{H}_{NL}}}
\def\cH{{\bf{H}}}
\def\cP{{\bf{P}}}
\def\cF{\bf{F}}
\def\cG{\bf{G}}
\def\bO{{\hbox{O}}}
\def\pN{\prod_{j=1}^N}
\def\sjN{\sum_{j=1}^N}
\def\skN{\sum_{k=1}^N}

\title{\bf On Non-Linear Quantum Mechanics\\[1in] and the Measurement Problem\\[1in] I. Blocking Cats\\[2in]}

\author{W. David Wick\footnote{email: wdavid.wick@gmail.com}}

\begin{document}
\maketitle
\pagebreak

\section*{Abstract}

Working entirely within the \Sch\ paradigm, meaning wavefunction only, I present a modification of his theory that prevents formation of states with macroscopic dispersion (MD; ``cats").
The proposal is to modify the Hamiltonian based on a method introduced by Steven Weinberg in 1989, as part of a program to test \qm\ at the atomic or nuclear level. By contrast,
the intent here is to eliminate MD without affecting the predictions of \qm\ at the microscopic scale. This restores classical physics at the macro level. 
Possible experimental tests are indicated and the differences from previous theories discussed.  
In a second paper, I will address the other 
difficulty of wavefunction physics without the statistical (Copenhagen) interpretation: how to explain random outcomes in experiments such as Stern-Gerlach, 
and whether a \Schist\ theory with a random component can violate Bell's inequality.
\pagebreak

\section{Introduction}
\subsection{The \MP\ and ``cats"}

In the 30-year-long debate over the interpretation of \qm\ (QM) by its founders and others, \Sch\ maintained that his 1926 formulation had been correct and moreover the \wf\ must contain an element of reality. 
He based that view on de Broglie's
insight that particles appeared to have wave properties; that his equation correctly predicted the spectrum of the hydrogen atom; and other observations \cite{smithsonian}, \cite{arethereqj}
such as ``G. B. Thomson's beautiful experiments on
the interference of de Broglie waves (of electrons), diffracted by crystals." He rejected the statistical interpretation of the wavefunction 
supported by Max Born, Niels Bohr, John von Neumann, Werner Heisenberg and others, as well as the individual identity of ``particles" which he likely believed are epiphenomena 
(appearing in consequence of the presence of a macroscopic measuring apparatus), as explained by Nevill Mott in 1929, \cite{mott}. 

However, in 1935 \cite{catpaper} \Sch\ described a difficulty with his theory when applied to measurement situations: it gives rise
to \md\ (MD) which he illustrated in the famous metaphor of a cat in a box containing a ``diabolical device": a 
bit of uranium, a Geiger counter, a hammer and a flask of cyanide, rendering the cat both alive and dead. To avoid life-and-death issues here, I revise the metaphor by replacing 
the hammer and poison by a buzzer; if it goes off, the cat is startled and leaps, say, one meter to the left. Thus the cat becomes smeared out between leaping and resting. 
I will call any similar situation appearing in quantum theory ``a cat".

Mathematically, we can describe MD as follows. 
Define the (spatial) dispersion of a macroscopic wavefunction to be:

\def\ooN{\left(\,\frac{1}{N}\,\right)}
\def\sN{\sum_{j=1}^N}

\bar
\no D_N(\psi^*,\psi) &\=& <\psi|\,\left\{\,\ooN\,\sN\,x_j\,\right\}^2\,|\psi> \-\\
&& \left\{\, <\psi|\,\ooN\,\sN\,x_j\,|\psi>\,\right\}^2.
\ear

\ni Here $N$ is the number of ``particles" or degrees of freedom, and  I have pretended that space has one dimension, 
purely for ease of writing equations; as there will never be a problem generalizing to three dimensions, 
I will maintain this pretense throughout the paper (except in the relativity section). 
A ``cat" will mean a \wf\ describing a macroscopic object with the peculiar property that $D_N$ is larger than its spatial extent (squared). 
Such cat states arise in \wf\ theory in measurement situations.

Consider the simplest such situation, that of measuring the ``spin of a single magnetic particle," which I assume takes only two values, $\pm 1$.
I write the corresponding ``spin states" as $|+>$ and $|->$. The measurement proceeds by sending the ``magnetic particle" 
through an inhomogeneous (external) magnetic field, which deflects ``the particle"
up or down by a certain amount in a specified time. Suppose the deflection is in the positive $y$-direction 
(while the ``particle" moves also in the $x$-direction, which isn't important for the discussion).
Thus an initial state of form

\be
\psi_0(y)\,|+>
\ee

\ni where $\psi_0(y)$ is a wave packet of small spread centered at the origin, evolves into another state:

\be
\psi_+(y)\,|+>
\ee

\ni with $\psi_+(y)$ displaced some distance in the $+$ $y$-direction. But this is not yet a measurement, because we have not yet included an apparatus capable of detecting ``the particle"
arriving up the $y$-axis. I will denote the apparatus location (say, its center-of-mass, COM) by $X$; think perhaps of a needle on a scale that can move in response to the ``particle's intrinsic magnetism."
So, initially, we should have written the combined state of everything as

\be
\psi_0(y)\,\theta_0(X)\,|+>
\ee

\ni where $\theta_0(X)$ is a wavepacket for the apparatus coordinates centered around the initial needle position,  and the state after the measurement as

\be
\psi_+(y)\,\theta_+(X)\,|+>
\ee

\ni where $\theta_+(X)$ is a wavepacket of the needle displaced by an amount we interpret as ``particle detected." (Better:  $\psi_{+}(y,x_1,...x_N)$, 
whose reduced density for $X$ is concentrated at the displaced position.) Similar results hold with $+$ spin replaced by $-$, with the needle moving down
rather than up.

So far this all seems quite satisfactory. But now the superposition principle produces a difficulty for \Schs\ theory.
Suppose the initial ``spin up" state is replaced by the superposition:

\be
(1/\sqrt{2})\,|+> \+ (1/\sqrt{2})\,e^{i\,\gamma}\,|->.
\ee

\ni (Here $\gamma$ is a real phase.) Then, by the unrestricted linearity of QM, the whole ``particle"-plus-measuring-apparatus state can only evolve into another superposition:

\be
(1/\sqrt{2})\,\psi_+(y)\,\theta_+(X)\,|+> \+ (1/\sqrt{2})\,e^{i\,\gamma}\,\psi_-(y)\,\theta_-(X)\,|->.
\ee

These developments of course represent simplified theory of the celebrated Stern-Gerlach experiment, performed in 1922,\cite{sandg}, 
in which the superposition was generated from an initial atomic
state of rotational symmetry. (Exact solutions of \Schseqn\ for S-G can be found in \cite{sandgmath1}, \cite{sandgmath2}.) 
It is presented in textbooks as the most distinctively ``quantum" outcome
and which refuted the classical picture of the atom as a little randomly-oriented magnet.
(I note here that the claim that Stern and Gerlach observed two dots is a myth;
it was actually a lip-print. What this may imply for measurement theory will be discussed in paper II of this series.) 

In my opinion, this scenario represents the fundamental conundrum arising in the QM theory from the 20th Century, often called the \MP. 
In the first place, no definite result has appeared, assuming we are unwilling
to believe that ``the particle" divided and went both ways.  
Second, as the needle is a macroscopic object, this version of measurement produces a \Schs\ (bidirectional) leaping cat. 
Finally, nothing has appeared to ``collapse the wavefunction" after the detection of the ``particle" up or down the y-axis. 

\subsection{What is the ``classical limit" of \qm?\label{class_section}}

Related to the Measurement Problem is the issue of the microscopic-to-macroscopic, quantum-to-classical, transition.
On the macroscopic side of the boundary, Newton's equations hold (at velocities small with respect to the speed of light); 
on the microscopic side, it is \Schs\ equation. What defines the boundary?
The conventional answer is: ``$\hbar \to 0$". However, I reject this opinion, for the following reasons. 

First: although Planck's constant certainly sets the energy levels of atoms and the magnitudes of many other phenomena in the quantum realm, it is a fixed constant of nature.
Taken literally, as an explanation of a classical realm, $\hbar \to 0$ is as absurd as claiming that the behavior of matter in a low-gravity region of interstellar space is due to a limit: 
$G \to 0$. Presumably what is meant is that  $\hbar$ is small in some sense. But in what sense, and how would that fix the classical/quantum boundary?
The units of $\hbar$ are [energy]x[time] or [momentum]x[distance]; 
thus, the statement ``$\hbar$ is small" does not in itself implicate a distance scale, a time scale,
an energy or momentum scale, or any combined scale that obviously distinguishes macroscopic from microscopic. 

Second, and somewhat more questionable, one can object based on the  existence of the phenomenon called ``chaos." 
Non-linear Newtonian equations in certain situations yield so-called ``chaos," and as one consequence
very complex orbital configurations in phase space may appear. By contrast, linear QM produces a quasi-periodic flow incapable of 
generating complexity of this sort \cite{fordandilg}. 
(Some physicists
reject this argument by invoking ``Heisenberg's uncertainty principle," which in pure \Schism\ we interpret simply as a theorem about forming wave packets.
In the usual probability interpretation of QM, this principle can be interpreted to mean that phase space is coarse-grained; i.e., observing a phase-space
trajectory with a precision below that permitted is impossible or meaningless. Therefore the chaos objection is invalid.)

Third: in my opinion, the most convincing reason to reject the $\hbar \to 0$ criterion is based on a calculation using
\Schseqn\ for an N-dimensional system that might represent an apparatus (see the Mathematical Appendix). Define
the center-of-mass (or centroid) of the system to be: 

\be
X \= <\psi|\,\left(\frac{1}{N}\right)\,\sum_{j=1}^N\,x_j\,|\psi>
\ee 

\ni and take two time-derivatives. There results:

\be
m\,\ddot{X} \= \- \left(\frac{1}{N}\right)\,\sum_{j=1}^N\,<\psi|V{'}(x_j)\,|\psi>.\label{ev_eqn}
\ee 

\ni where $V$ denotes an external (scalar) potential; a dot denotes time-derivative; and a superscript $'$ denotes space-derivative. 

Note two interesting facts about this equation. One: $\hbar$ dropped out. Two: suppose that in the right-hand side the average over $j$ and the integrals over space can be pushed inside
the function $V'$, up to some negligible error. The result is Newton's equation for the COM:

\be
M\,\ddot{X} \= \- N\,V{'}(X) \equiv - V'_E(X).
\ee 

\ni ($M \equiv N\,m$ and the factor of $N$ on the right 
is absorbed into a ``macroscopic" external potential energy,
rendering it extensive.) 
In particular, for harmonic potentials, where $V'$ is linear, Newton's law holds without qualifications, as many authors have noted previously. I conclude from this calculation that we 
should search for the classical limit where {\em the dispersion of the COM of the apparatus pointer is very small relative to the scale on which the higher-order terms in
the external potential vary spatially.}

\subsection{Resolving the paradox of measurement\label{resolving_section}}

One way to describe the MP is to note that QM produces an `and'---the wavepackets of S-G propagate up {\em and} down---when what we observe is an `or': a detection up {\em or} down. 
Rather esoteric proposals have been made to resolve this paradox: the mind of the observer collapsing the wavefunction (von Neumann, 1932, \cite{vnbook}); 
the universe splitting into branches in the ``multiverse"
(Everett 1957, \cite{everett}), and many others. However, if macroscopic dispersion can be eliminated at the level of apparatus, 
the `and' must necessarily be transformed into the `or'. 
Extra-physical postulates will no longer be required to explain measurements. That is the motivation for the theory proposed in the subsequent sections.

\section{\NLQM\label{nlqm_sect}}

\subsection{Weinberg's formalism\label{weinform_section}}

I will adopt Weinberg's approach, \cite{weinberg}, to defining a class of nonlinear, deterministic evolution equations for a complex wavefunction $\psi$, 
which is simple and elegant. (Weinberg's results lead to experimental tests of whether QM has non-linear contributions at the nuclear level, with negative results; see \cite{weinbergtests}.)
The starting point is a minor
reformulation of \Schs\ theory, essentially a change of language. Traditionally the Hamiltonian, or energy observable, is thought of as a self-adjoint operator, written $H$, and \Schs\
equation takes the form:    

\be
i\,\hbar\,{\pa\,\psi\,/\,\pa\,t} \= H\,\psi.
\ee

Instead, from now on I will use ``Hamiltonian" to mean the functional $\cHqm$ = $<\psi|H|\psi>$. (Which in the statistical interpretation of QM is called the ``expected energy." 
But \Schists\ 
have abandoned that probabilitistic conception.) We rewrite \Schseqn\ as: 

\be
i\,\hbar\,{\pa\,\psi\,/\,\pa\,t} \= \pa\,/\pa \psi^*\,\cHqm.
\ee

Note that here we regard the \Ham\ to be a functional of $\psi$ and $\psi^*$ rather than the real and imaginary parts of $\psi$, but the two views are equivalent.
(There is no problem about varying $\psi$ and $\psi^*$ independently, as in the usual definition of partial derivative; I define it as if by the chain-rule, see Mathematical Appendix.)
 Now Weinberg's idea is simply that,
given any other real-valued function $\cH(\psi*,\psi)$, which need not be quadratic in the $\psi$-variables, we can postulate a new equation:  

\be
i\,\hbar\,{\pa\,\psi\,/\,\pa\,t} \= \pa\,/\pa \psi^*\,\cH.\label{geneq}
\ee

For its application to the MP I will assume an additive form:

\be
\cH \= \cHqm \+ \cHnl.\label{ham}
\ee

\ni where $\cHnl$ will be small in a sense to be discussed later.

It is easy to check (see \MA) by separating real and imaginary parts that (\ref{geneq}) 
is simply a way of packaging Hamiltonian mechanics, albeit for a pair of ``position" 
and ``momentum" variables which are not the physical quantities. All of the Hamiltonian machinery goes through; for instance, the Poisson bracket takes the form, 
for any pair of functionals $\cF(\psi^*,\psi)$
and $\cG(\psi^*,\psi)$:

\be
\left\{\,\cF,\cG\,\right\}  \= <\,\frac{\pa\,\cF}{\pa\,\psi}\,\frac{\pa\,\cG}{\pa\,\psi^*} \- \frac{\pa\,\cF}{\pa\,\psi^*}\,\frac{\pa\,\cG}{\pa\,\psi}\,>.
\ee

\ni which reduces, in the case that  both functionals are quadratic, defined by two self-adjoint operators $F,G$, to the usual ``commutator" form:

\be
\left\{\,\cF,\cG\,\right\}  \= <\psi|\,\left[\,F,G\right]\,|\psi^*>.
\ee

\ni where $\left[\,F,G\right]$ = $F\,G - G\,F$. In terms of this bracket, the time-derivative under the evolution defined by equation (\ref{geneq}) is given by:

\be
d\,/\,dt\,\cF \= \left(\,i\,\hbar\,\right)^{-1}\,\left\{\,\cF,\cH\,\right\}.
\ee

\ni In particular, energy is conserved:

\be
d\,/\,dt\,\cH \= \left(\,i\,\hbar\,\right)^{-1}\left\{\,\cH,\cH\,\right\}  \= 0.
\ee

What about conservation of norm, interpreted in the statistical paradigm as ``conservation of probability?" (A \Schist\ might not care; on the other hand, we could
worry about conservation of matter or charge.) In nonlinear theory we have:

\be
\frac{\pa}{\pa\,t}\,<\psi|\psi> \= \frac{-i}{\hbar}\,\left\{<\psi|\,\frac{\pa\,\cH}{\pa\,\psi^*}> - <\frac{\pa\,\cH}{\pa\,\psi*}|\psi>\,\right\},
\ee

\ni which is not obviously zero. It is in special cases, in which 

\be
\frac{\pa\,\cH}{\pa\,\psi^*} \= O(x_1,...;\psi)\,\psi,
\ee

\ni and $O(x_1,...;\psi)$ happens to be a self-adjoint operator possibly depending on $\psi$ in some fashion. (Weinberg apparently 
rejected this approach. Hence, in order to remain consistent to the statistical interpretation, he had to divide
by the normalization when discussing ``expected values" in his nonlinear models.) That will be the case here; see next section and the \MA.

\section{Blocking Cats\label{blocking_section}}

To motivate the introduction of nonlinear terms into \Schseqn, consider first two wavefunctions that can represent states of a macroscopic system.   
Let $\phi_r(x)$ be a wavefunction mostly confined to an interval $[-r,r]$ with $r$ fairly small, say a few centimeters.
Let $R$ denote a larger distance, say a few meters. Compare:

\def\oos{\left(\,1/\sqrt{2}\,\right)}

\be
\pN\,\left\{\,\oos\,\phi_r(x_j + R) \+ \oos\,\phi_r(x_j - R)\,\right\}.\label{wvone}
\ee

with:

\be
\oos\,\pN\,\phi_r(x_j + R) \+ \oos\, \pN\,\phi_r(x_j - R).\label{wvtwo}
\ee

Wavefunction (\ref{wvone}) can represent a rather peculiar extended object with ``particles appearing at two locations" (simultaneously), a situation that can arise with so-called \MQP, 
see section \ref{exp_sect}. (If I had left out the superpositions and simply replaced $\phi_r$ by a $\phi_R$ with width $2\,R$, we probably would not worry about this issue.) But it is not a cat.
Wavefunction (\ref{wvtwo}) is a (bidirectional) leaping cat. How can we distinguish the two, mathematically?

Here we rely on the dispersion $D_N$ introduced earlier. It is easy to see that, for the non-cat extended state of (\ref{wvone}), $D_N = \bO(R^2/N)$, 
while for the leaping cat state in (\ref{wvtwo}), $D_N = \bO(R^2)$,
differing by a factor of $N$ (perhaps \pluspp{23}\ ). 
(If we believed the statistical interpretation of the \wf, this is the statement of the Central Limit Theorem.
For \Schists, it is a fact about a mass distribution with independent contributions.) 

I therefore propose to exploit conservation of energy in a nonlinear \Schseqn\ to prevent the formation of the cat. Let

\bar
\no \cHnl(\psi^*,\psi) &\=& w\,\left\{\, <\psi|\,\left(\,\sN\,x_j\,\right)^2\,|\psi> 
 \- \left(\, <\psi|\,\sN\,x_j\,|\psi>\,\right)^2\,\right\}\\
 &\=& w\,N^2\,D_N(\psi^*,\psi).\label{nlsch_eqn}
\ear 

\ni Here $w$ is a (very) small coupling constant with units [energy]/[distance]$^2$.
(Alternatively, one can construct a momentum version of $\cHnl$ for which $w$ has different units; see section \ref{rel_sect}.)

Adding $\cHnl$ to the usual ``N-particle" quantum Hamiltonian $\cHqm$ generalized from equation (\ref{ham}), the latter perhaps including internal energy terms holding the apparatus together,
external fields, and 
connection to a microsystem,
yields my proposed NLQM. (Another difference with Weinberg appears here: Weinberg assumed that all observables must be homogeneous of degree one in $\psi$ and $\psi^*$, because,  
if $\psi$ is a solution of a nonlinear \Schseqn, he required that  $Z\,\psi$ be also, for any complex number $Z$. 
I require only that this property hold for complex numbers with $|Z| = 1$, i.e., which maintain \wf\ normalization.)

The key to this proposal is that, if cats exist, $\cHnl$ does not scale ``extensively," proportionally to $N$, as is assumed with other kinds of energy.
If we suppose that, in a measurement situation where the apparatus is coupled to a microsystem, the initial state is of form:

\be
\pN\,\phi_r(x_j)\,\theta(y)
\ee

\ni (where $y$ stands for some microsystem coordinates), then the initial energy consists of apparatus internal and external energies (scaling like $N$), a microscopic energy
from coupling to the microsystem, plus a contribution from the NL term of order $w\,r^2\,N$.
However the total system evolves, it cannot
become a leaping cat, because there is not enough energy available to reach $\bO(w\,R^2\,N^2)$ (provided $w$ is not too small). 

For illustrative purposes, I invent a value for $w$ and see how the numbers come out. 
Assuming that all apparatus energies are positive (or at least bounded below by something times $N$),
the final nonlinear energy term after the measurement cannot exceed the initial energy. But if the measurement produces a leaping cat, the nonlinear term will be of
order

\be
\cHnl[\hbox{final}] \siml w\, [1m]^2\,10^{46}.
\ee

If we postulate

\be
w \siml 10^{-30}\, \hbox{Joule}/m^2,
\ee

Then 

\be
\cHnl [\hbox{final}]\siml 10^{16} \,\hbox{Joule}.\label{lpc}
\ee

However, if the apparatus consists of a pointer
of size $1$mm (plus external fields) initially

\be
\cHnl[\hbox{initial}] \siml 10^{-30} \,10^{23}\,10^{-6}\,\hbox{Joule} \siml 10^{-13}\,\hbox{Joule}.
\ee

Clearly, the energy of the leaping cat in (\ref{lpc}) cannot be supplied by anything (apparatus internal energy, external fields, the microsystem) in this situation. 
For perspective: if the cat weighed 1kg and the potential energy of (\ref{lpc}) were converted into kinetic energy (ignoring relativistic mass increase), 
the cat would be travelling at around \plusp8\ meters per second, 
i.e., near the speed of light.

By contrast, the nonlinear energy of a hydrogen atom would be around

\be 
10^{-30}\,\left[6\times10^{-11}\right]^2 \sim 10^{-51}\, \hbox{Joules},
\ee

\ni which is about \negpp{32}\ times the ground state energy ($1.6\times10^{-19}$ Joules), a level which could never be observed.

\subsection{Restoring Newton\label{restoring_sect}}

One of the requirements for claiming to have a solution of the MP is to restore Newton's Laws at the macroscopic side of the classical/quantum boundary.
How would this work in the NLQM proposed in the previous section? 

First, let us futher examine how an energy bound on the nonlinear terms controls the mass distribution of the apparatus. For concreteness, 
visualize a pointer suspended initially at an unstable extrema of an applied,
external, potential; the pointer will move upon absorbing energy from the microsystem observed. I will need the reduced density functions of the pointer. 
The $k$--th reduced density function is defined as: 

\be 
\rho_k(x_1,...,x_k) \= \int ... \int\, dx_{k+1}\,dx_{k+2} ... dx_N\,|\psi|^2(x_1,...,x_N).
\ee

(I assume symmetry under interchange of variables.) 
In terms of reduced functions the macroscopic dispersion can be written: 

\bar
\no D_N &\=& \left[\frac{N\,(N-1)}{N^2}\right]\, \left\{\,\int\int\,\rho_2(x_1,x_2)\,x_1\,x_2\,dx_1\,dx_2 \- \left(\,\int \rho_1(x)\,x\,dx\,\right)^2\,\right\}\\
\no && \+ \left[\frac{1}{N}\right]\, \left\{\,\int\,\rho_1(x)\,x^2\,dx \- \left(\,\int\,\rho_1(x)\,x\,dx\,\right)^2\,\right\}\\
&& \= \left[1 - \frac{1}{N}\right]\,C_2 \+ \left[\frac{1}{N}\right]\,L^2.
\ear

\ni where the third line defines the ``correlation function" 
$C_2$ and the squared size of the system, $L^2$. 
The nonlinear energy becomes, omitting a small term:

\be
\cHnl \= w\,N^2\,D_N \sim w\,\left\{N^2\,C_2 + N\,L^2\right\}.
\ee

Thus an energy bound on $\cHnl$ implies a strong bound 
on $C_2$ but a weaker bound on $L^2$. 

However, to transition to Newton, a bound on $L^2$ is crucial. 
Consider a cubic external potential:

\be
V_E(x) \= a\,x \+ b\,x^2 \+ c\,x^3.
\ee

By an easy computation using (\ref{ev_eqn}),

\be
M\,\ddot{X} \= \- V_E{'}(X) \- 3\,c\,L^2.\label{xeqn}
\ee 

\ni where $X$ is the pointer COM and I have omitted the interaction term with the microsystem, which can be thought of here as simply giving the pointer an initial (small) kick. 
Thus Newton for the pointer is restored as long as the second term on the right of (\ref{xeqn}) is small, which will be true if the size of the pointer remains smaller
than the variation scale of the cubic term. I note that no additional terms coming from $\cHnl$ appear in this computation, because adding $\cHnl$ does not affect the derivation of 
(\ref{ev_eqn}), the macroscopic evolution equation. (See \MA.
Of course, a cubic is not an ideal example, especially for the energy bound, as it is not bounded below.)

Now let us assume that the observation does not produce a phase transition (i.e., the pointer doesn't melt or vaporize) or an explosion.
Moreover, the pointer is held together by internal forces, say corresponding to pair potentials with energy:

\be
\- \sum_{j\neq k}^N\,<\psi|\,u(x_j - x_k)\,|\psi>,\label{cubic_case}
\ee

\ni with $u(x)$ becoming small as soon as $|x|$ is larger than some microscopic length, say a few lattice spacings. Then the internal 
(negative) binding energy will be extensive (proportional to $N$) and much larger than energy supplied from the observed microsystem. 
Now $L^2$ is the squared average distance between atoms and the centroid of the pointer; if it increased, atoms would be farther apart, and the internal energy would increase proportionally. 

I postulate, as part of measurement theory within \Schism, that $X$, $D_N$ and $L^2$ are observables, and ``restoring Newton" for the apparatus pointer means that $X$ evolves by his equation while
$D_N$ remains very small and $L^2$ does not increase. The energy bound on $\cHnl$ ensures that $D_N$ remains small, 
while internal energies ensure that $L^2$ doesn't increase. 
Then, at least for the cubic case, (\ref{xeqn}) yields Newton. 

What about higher-order terms in the external potential? (A quartic would be more realistic.) I do not have a theorem here, but I postulate that the tail thickness of
realistic examples of the reduced density function $\rho_1$ are controlled by the second moment, as for Gaussians, 
implying that higher-order terms in (\ref{xeqn}) are also small. 

\subsection{Other properties of NLQM\label{spec_sect}}

I remarked that energy is conserved in the nonlinear extension of QM; what about momentum? Define the total momentum of a many-body system to be:

\be
\cP \= <\psi|\,\sjN\, \frac{i\,\hbar\,\pa}{\pa\,x_j}\,|\psi>,
\ee

Then (easy exercise):

\be
\left\{\,\cP, \cHnl\,\right\} \= 0.
\ee

It follows that if $\cHqm$ contains only internal forces (no external fields) plus the usual momentum terms,

\be
\left\{\,\cP, \cH\,\right\} \= 0,
\ee

\ni so total momentum is also conserved. (Apparently Weinberg adopted a strict homogeneity condition in the nonlinear Hamiltonians
he considered because it was necessary to ensure Galilean invariance. As $\cHnl$ is obviously
invariant under overall rotations, space translations, and boosts, I see no difficulty here. Lorenz invariance is another matter.) 

How does the NL energy behave for a composite system? If the two subsystems are independent, in the sense that they have never interacted, or never interacted with the same other system,
and the \wf\ factorizes, then this energy is additive: if the subsystems are labeled A and B,

\def\cHnlA{{\bf{H}_{NL;A}}}
\def\cHnlB{{\bf{H}_{NL;B}}}
\be
\cHnl \= \cHnlA \+ \cHnlB.
\ee

In particular, if

\be
\psi \= \pN\,\theta_r(x_{j;A} + R)\,\prod_{k=1}^N\,\theta_r(x_{k;B} - R),
\ee

\ni where $\theta_r(\cdot)$ is a wavepacket of width $r$ centered at the origin, then $\cHnl$ is proportional to $r^2$, not $R^2$. So this peculiar form of energy cannot be thought of
as a ``confining potential" (as for quarks I suppose) preventing the separation of the two subsystems. 
For independent systems combined mentally into one larger system, $\cHnl$ scales extensively, as usual. 

If macroscopic subsystems A and B would become correlated by interacting with a third (microscopic) system, 
the NL energy kicks in and prevents cat-formation as usual, except for one purely-theoretical instance:
EPRB with perfect detectors, in the perfectly-anticorrelated case. (EPRB is discussed in the second paper of this series.) 
If the detectors are needles that move in opposite directions, the dispersion of the COM of the combined system could remain zero, abrogating the cat-blocking mechanism proposed here.
However, this case requires exact alignment ($a = b$), identical detectors, and identical initial conditions. With any discrepancy the nonlinear terms would scale again as $N^2\,R^2$ and energy conservation
would prevent cats. Additionally: perfect anticorrelations are a consequence of assuming a perfect von Neumann measurement (in fact, Stern and Gerlach did not observe two dots; see paper II)
and may be unrealistic.

Thus far, I have only considered two macrosystem wavefunctions: a product form (``independent particles") and the cat. 
For the only ``normal" state (the former), the quantity $C_2$ is exactly zero. But for most states of matter, it will not vanish.
This raises the issue of ``correlations" in \Schism. 

First, $C_2$ should not be called a ``correlation function," as that language is only suitable for the 
statistical interpretation of the \wf\ (as representing, in some complicated way, an ensemble of particles). 
$C_2$ can only be regarded as an aspect or component of the dispersion of the system. 
Second, to construct a \Schist\ version of ``quantum statistical mechanics," or ``quantum thermodynamics," 
the only consistent approach is to introduce an ensemble of wavefunctions, perhaps describing the system interacting with a heat bath at temperature $T$.
But there is little agreement in the literature on how this is done. Some authors adopt {\em a priori}
an ensemble weighted by Gibbs factors: $\exp(\,- E/kT\,)$, where $E$ is what I have written $\cH$ 
and $k$ is Boltzmann's constant.
Justifying such an assumption, whether the dynamics is classical or quantum, is famously hard. 
Then there is \vN's conception of entropy,
defined in terms of a ``density operator" usually written $\rho$ (not a reduced density function as I used earlier), 
which subsumes both ``pure states" (single wavefunctions) and ``mixtures" (as in probability theory).
It takes the form:

\be
 \rho \= \exp(-H/kT)/\hbox{tr}\exp(-H/kT).
\ee

\ni (``tr" stands for matrix trace.) Then one defines the entropy as: 

\be
 - k\,\hbox{tr}\,\rho\,\log(\rho).
\ee

 Unfortunately, under linear (``unitary") matrix evolution using Heisenberg's rule:

\be
\frac{d}{dt}\,\rho \= \frac{-i}{\hbar}\,[\rho,H],
\ee

\ni such an entropy is constant! This violates the Second Law of Thermodynamics, at least as formulated: ``outside of equilibrium, entropy always increases". Many authors
have addressed this problem; some introduce nonlinear, dissipative evolution to restore the Second Law. 
It has also been proposed that statistical mechanics and thermodynamical behavior can be derived from linear QM 
either because the system has been subjected to a random Hamiltonian \cite{Deutsch} (I make a similar suggestion in paper II to explain the random outcomes of certain measurements), 
or because the classical analog of the system is chaotic, \cite{Sred}; see \cite{PandS} for a recent review of this program. 
But I find no consensus about the proper formulation of ``quantum statistical mechanics."  

Third, thermodynamic equilibrium reflects a balance of energy and entropy. Thus introduction of a new form of energy 
that involves correlations (in some sense), which are usually thought of as an aspect of entropy, may alter the nature of phase transitions or the location of critical points.
(Before this topic can be addressed, it will be necessary for experiments to fix the free parameter called ``$w$.") However, as phase transitions do not implicate cats, this is unlikely.
Finally, even having adopted a thermodynamic theory, we receive no guidance as to the form
of $C_2$ for an individual \wf. 

I make a few conjectures about the form of $C_2$ for more general cases. Writing:

\be
\frac{\rho_2(x_1,x_2)}{\rho_1(x_2)} \= \rho_1(x_1)\,\left[\,1 + f(x_1-x_2)\,\right],
\ee

\ni we can think of $f(x_1 - x_2)$ as representing the fractional increase or decrease in the one-dimensional (``single particle") density at $x_1$, conditional on ``a particle at $x_2$." 
We can imagine that $f(x)$ falls falls off rapidly with $|x|$, say over a microscopic distance. 
Then we have:

\be
C_2 \= \int\int\,dx_1\,dx_2\,\rho_1(x_1)\,\rho_1(x_2)\,f(x_1 - x_2)\,x_1\,x_2,
\ee

\ni (where again I have written this equation as though space has one dimension). I worked out a simple example (see the Mathematical Appendix) and found

\def\maxf{\hbox{max}(f)}.
\be
C_2 \approx L\,\epsilon\,\maxf.
\ee

\ni where $L$ denotes the size of the system and $\epsilon$ is a microscopic length (the ``correlation length"). Now the issue becomes: how large can $\maxf$ be? 
It represents the maximal increase (or decrease) in density at some point, due to knowledge that there is ``a particle nearby." If we assume that this is small, of order $N^{-1}$,
then we have the usual (``Central Limit") behavior; this would result in a contribution of the nonlinear energy for a ``normal" object that is very small. 
On the other hand, if $\maxf \sim 1$, then with my illustrative value for $w$, the $1$mm needle, and nanoscale correlations, we get a contribution of order:

\bar
\no w\,N^2\,L\,\epsilon &\approx& 10^{-30}\,10^{46}\,10^{-3}\,10^{-9}\\
\no &\approx& 10^4\,\hbox{Joules}.
\ear

\ni This is roughly, for example, the total thermal energy of a 0.1g mass with the heat capacity of water, at room temperature. 
That is large; but I suspect that, in realistic models of pointers, what I
have written here as $\maxf$ is small enough that the contribution to the energy of the object is very small.

\section{Relativistic generalizations\label{rel_sect}}

Can a theory of the type considered here be compatible with the Relativity Principle? 
Although modern \Schists\ need not follow the historical path, 
it is useful to recall the order of 20th Century developments in physics. After Heisenberg and \Sch\ published in 1925-6, it must have seemed unlikely that the new quantum mechanics 
could ever be made compatible with the RP. 
Then, in 1928, Paul Dirac found his miraculous equation, which was.
But the full union of QM with relativity was supposedly attained only in the Quantum Field Theory, 
a theory of ``infinitely many particles" dating from the 
1940s.

Can we construct a relativistic \wf\ generalization of the nonlinear theory in section \ref{nlqm_sect}?
The best approach may be to start from the momentum form of the nonlinear Hamiltonian.  
Reasoning that in order for the cat to leap she must 
acquire a momentum, I could have defined the nonlinear term $\cHnl$ by replacing $x_k$ by $p_k = i\,\hbar\,\pa/\pa x_k$, the momentum operator, in 
equation (\ref{nlsch_eqn}):

\bar
\no \cHnl(\psi^*,\psi) &\=& w\,\left\{\, <\psi|\,\left(\,\skN\,p_k\,\right)^2\,|\psi> \-\right. \\
\no&&\left. \left(\, <\psi|\,\skN\,p_k\,|\psi>\,\right)^2\,\right\}.\\ \label{nltermsmomentum}
\ear

\ni Here the parameter
$w$  has units of reciprocal mass; alternatively, if I incorporate such a factor explicitly, $w$ is unit free. (Neither choice seems as felicitious for locating the classical/quantum boundary
as the spatial formulation. 
But either choice does the job: blocking cats.)

The relativity problem with the \Schseqn\ was that it includes the energy operator ($i\,\hbar\, \partial/\partial\, t$) on the left side but the {\em square} of the momentum operator on the right;
relativity requires that energy and momentum be treated equivalently, as components of the ``energy-momentum 4-vector." 
In 1926 Oskar Klein and Walter Gordon (and many others)
proposed a new equation similar to a classical wave equation:

\be
\frac{1}{c^2}\,\frac{\partial^2\,\psi}{\partial\,t^2} \= \triangle\,\psi \- \frac{m^2\,c^2}{\hbar^2}\,\psi.
\ee

The obvious generalization of the Klein-Gordon equation to N ``particles" and including the nonlinear energy is to write:

\be
\frac{\hbar^2}{m\,c^2}\,\frac{\partial^2\,\psi}{\partial\,t^2} \= \frac{\partial}{\partial\,\psi^*}\,\left\{\,\cHqm + \cHnl\,\right\} - N\,m\,c^2\,\psi.\label{nlkgeqn}
\ee

\ni where now

\be
\cHqm \= \frac{1}{2\,m}\,\skN\,<\psi|\,p_k^2\,|\psi>
\ee

\ni is the usual total kinetic energy and $\psi$ remains as a (scalar) complex function of all the ``particle" coordinates and time. 

Although K and G proposed their equation as representing a relativistic electron, Dirac rejected it, for two reasons. (More recently, K-G seems to 
have been resurrected as the correct equation to describe the pion and the Higgs boson.) First, it did not incorporate 
the intrinsic ``spin" of an electron. And, as a second-order equation, two initial values could be arbitrarily prescribed: 
for $\psi(x_1,...,0)$ and $\partial\,/\partial\,t\,\psi(x_1,...,0)$, and so normalization and the 
statistical interpretation of the \wf\ could not be preserved. For the multiparticle case, there is another objection (that may not have been raised at the time): 
KG is incompatible with the notion of ``independent particles". 
Given a product form for, e.g., two particles: $\psi_1(x_1;t) \psi_2(x_2;t)$, even if each factor satisfies a one-particle KG equation, because of the two time derivatives
on the left side, the product will not satisfy a two-particle KG equation. Assuming the product form at time zero, it will not propagate. 
Thus, even lacking external fields or any other interactions, particles would become entangled.

Dirac's clever solution to these dilemmas was to increase the number of components of $\psi$ to four,
and then to factorize K-G by matrix tricks. His equation, using the celebrated gamma-matrices, reads (for this section I restore space to its rightful number of dimensions, namely three):

\def\gammamu{\gamma^{\mu}}
\def\parmu{\partial_{\mu}}
\bar
\no \gamma^{\mu}\,\left\{\,p_{\mu}\, + \frac{e}{c}\,\Phi_{\mu}\, \right\}\,\psi \- m\,c\,\psi &\=& 0;\ph\hbox{or:}\\
\gamma^{\mu}\,\left\{\,i\,\hbar\,\partial_{\mu}\, + \frac{e}{c}\,\Phi_{\mu}\, \right\}\,\psi \- m\,c\,\psi &\=& 0. \label{dirac_eqns}
\ear

\ni Here $\mu = 0,1,2,3$ indexes space-time components ($0$ is for time), with a repeated Greek index implying summation; $\gammamu$ are 4x4 matricies satisfying the anti-commutation relations:

\be
\no \gamma^{\mu}\,\gamma^{\nu} \+ \gamma^{\nu}\,\gamma^{\mu} \= 2\,g^{\mu,\nu}\,I;
\ee

\ni $g^{\mu,\nu}$ denotes the space-time metric, diagonal$(1,1,1,-1)$; $\parmu = (\frac{1}{c}\,\partial/\partial\,t,\partial/\partial\,x,\partial/\partial\,y,\partial/\partial\,z)$;
$\Phi_{\mu}$ is the 4-potential of the external electromagnetic field; $m$ is the electron mass; $c$ is the speed of light; and $\psi = (\psi_0,\psi_1,\psi_2,\psi_3)$.

An obvious approach is to treat the momentum terms in (\ref{nlkgeqn}) as Dirac did, granted a sufficient supply of anti-commuting $\gamma$-matrices to factorize them. 
For Lorenz covariance
of either a nonlinear K-G or Dirac equation we would also need to know that:

\be
<\psi'|P'_{\mu}|\psi'> \= \Lambda^{\nu}_{\mu}\,<\psi|P_{\nu}|\psi>,\label{transeq}
\ee

\ni where primes denote a second reference frame; $P$ and $P'$ stand for momentum operators; $\Lambda^{\nu}_{\mu}$ is the matrix of a Lorenz transformation, denoted $\Lambda$, connecting the frames;
and, in the Dirac case, in terms of an invertible 4x4 matrix $S$ depending on $\Lambda$:

\be
\psi'(x_1',...) \= S(\Lambda)\,\psi(\Lambda(x_1,...)).\label{trans_eqn}
\ee

\ni In words: energy-momentum transforms 
as a covariant 4-vector. (In the Dirac case it is necessary to define in equation \ref{transeq}:

\be
<\psi|\cdot |\psi> \= \int\,d^3x\,\psi^{\dagger *}\beta\,[\cdot ]\psi
\ee

\ni where $\beta = i\,\gamma^{0}$, because the matrix $S(\Lambda)$ is not unitary but ``pseudounitary": $\beta\,S(\Lambda)^{\dagger *}\,\beta = S(\Lambda)^{-1}$, see \cite{WeinbergQFT}, p. 218. 
The probability interpretation for the electron's position is then lost, because $\beta$ has eigenvalues $\pm 1$; but \Schists\ do not require that interpretation to hold.)

So let us assume given a set of matrices acting on some vector space in which $\psi$ takes values and satisfying:

\be
\no \gamma^{\mu}_j\,\gamma^{\nu}_k \+ \gamma^{\nu}_k\,\gamma^{\mu}_j \= 2\,\delta_{j,k}\,g^{\mu,\nu}\,I,
\ee

\ni (clearly such a set of generalized gamma-matrices requires a high-dimensional vector space on which the matrices act, 
so we might as well have assumed an infinite-dimensional
$\psi$ at the outset),
\ni for $j,k = 1,\dots N$, and define

\be
\gamma^{\mu} \= \sum_{j=1}^N\,\gamma^{\mu}_j.
\ee

\ni We can then postulate (omitting the external potentials for simplicity):

\be
\sum_{k,\mu}\,\gamma^{\mu}_k\,i\,\hbar\,\partial_{\mu,k}\,\psi + w\,\sum_{\mu}\,\gamma^{\mu}\,\left(\,\sum_k\,\left[\,i\,\hbar\, \partial_{\mu,k} - 2\,<\psi|P_{\mu,k}|\psi>\,\right]\right)\,\psi 
= \sqrt{2\,N}\,m\,c\,\psi.
\ee

Next, ``squaring" the equation (applying the operator on the left a second time), using the anticommutation relations, and ignoring terms of order $w^2$ yields:

\bar
\no && \- \sum_{k,\mu}\,g^{\mu,\mu}\,\hbar^2\,\partial^2_{\mu,k}\,\psi \+ w\,\sum_{\mu}\,g^{\mu,\mu}\,\left\{\,\sum_{j}\,i\,\hbar\,\partial_{\mu,_j}\,\right\}^2\,\psi\\
\no && \- 2\,w\,\sum_{\mu,j,k}\,g^{\mu,\mu}\,i\,\hbar\,\partial_{\mu,j}\,<\psi|P_{\mu,k}|\psi>\,\psi\\
&& = N\,m^2\,c^2\,\psi.
\ear

\ni Here the first term represents the usual total KE and the second and third yield a nonlinear KG equation similar to (\ref{nlkgeqn}) but with additional centering term for the energy 
(from terms with $\mu = 0$). 
The mutual repulsions of electrons could be incorporated through external potentials as in (\ref{dirac_eqns}), using the Dirac current as source terms for Maxwell's equations.


It is well known that Dirac's theory suffered from anomalies. The existence of states with negative kinetic energies was the primary difficulty; Dirac proposed re-interpreting these states as
representing positrons (negative electrons). But then external potentials might induce transitions from electron to positron states or {\em vice versa}, violating conservation of charge;
so Dirac assumed the negative-energy states were all filled. There are difficulties with observables. The usual position observable 
mixes the two kind of states, and the usual velocity operator produces a peculiar motion (the Zitterbewegung). 
Restricting to positive states or introducing some novel position operator leads to a violation of Einstein causality 
(wave packets spread faster than the speed of light).\cite{thaller}
R. Jost concluded: ``The unquantized Dirac field has 
therefore no useful physical interpretation."\cite{jost}

These conundrums suggest that it may be a mistake to start the search for a relativistic, nonlinear \Schist\
theory with Dirac. Should we follow the historical sequence and progress to Quantum Field Theory, even if it is considered the quintessential theory of particles?
QFT employs ``creation and annihilation" operators, usually described as creating or destroying particles; 
but they could alternatively be interpreted as amplifying or simplifying wavefunctions 
(thus making $N$ another argument). 
Thus QFT, if reinterpreted, may be compatible with \Schism. On the other hand, A. O. Barut and colleagues showed in the 1980s (see \cite{barut} and references therein) 
that many successful calculations in QFT---including of the Lamb shift, spontaneous emmission, vacuum polarization, 
and the anomalous magnetic moment of the electron---can also be carried out in a purely \Schist\ setting. 
A nonlinear Dirac-type equation, derived from integrating out the Maxwellian potentials, appeared in their program (but I am not aware that they related it to the Measurement Problem).

\section{Experimental Tests\label{exp_sect}}

\def\dV{\Delta\,V} 
What do we need to test the theory presented here? 
Before describing several candidate systems, let's ask: have observations already falsified the theory?
Certain systems studied by quantum physicists are said to exhibit ``macroscopic quantum phenomena" (MQP). Do these phenomena include MD (cats)?
Back in 1980,\cite{leggett1980}, Anthony Leggett, 
after discussing cats etc., had this to say: 

\bq

To sum up the point crudely and schematically, ``macroscopic quantum phenomena" require a many-particle wave function of the form,

\be
\psi \= \left(\,a\,\phi_1 \+ b\,\phi_2\,\right)^N,
\ee

\ni while the states of importance in the quantum theory of measurement are of form

\be
\psi \= a\,\phi_1^N \+ b\,\phi_2^N.
\ee
\eq

\ni As we have seen, the second equation represents a cat, but not the first. Thus MQP is not (necessarily) relevant to the MP.

In addition, systems exhibiting MQP are very complex, require cooling to extremely low temperatures, and theory describing them often relies on heuristic or uncontrolled assumptions.  
Leggett has stated that the exemplar of MQP is the Superconducting Quantum Interference Device (SQUID), consisting of a superconducting ring with
one or more Josephson junctions. This device is almost macroscopic (it can be seen with the naked eye). However, the theory of the device is based on modeling the magnetic flux through the ring as if it
were a one-dimensional quantum variable, with its own \Schs\ equation. \cite{leggett2002}
Moreover, Leggett has stated \cite{leggett2000}
that while SQUIDS exhibit quantum phenomena such as entanglement and interference, the number of electrons in the circuit  
is only \plusp9--\pluspp10; I understand that as telling us that SQUIDS 
lie on the quantum side of the classical/quantum boundary.

For a test, we need a scalable quantum/classical system---one that exhibits quantum phenomena at the microscale but transitions to classical behavior at larger size, 
keeping all else fixed. Given such an experimental set-up, testing would proceed in two stages. 
First, the elimination of macroscopic center-of-mass (COM) dispersion (``cats") must be used to fix the free parameter I have denoted by ``$w$". 
Then the now-rigid theory should predict, say, the shape of the transition curve, or the disappearance of some other quantum phenomena (such as interference). 

Unfortunately, 
the SQUID is not suitable as it does not produce a separation of the COM of the electrons in the circuit, \cite{leggett2002}.
A more promising choice may be the ``micromechanical oscillator": a nearly-atomic-sized ``beam" or bridge fixed at both ends and observed at low temperatures, 
for which quantum phenomena such as discrete energy states have been observed, \cite{micromech}. A superposition of macroscopic motional states may be possible soon in such a system, \cite{vienesse}, 
although achieving large displacements (greater than the size of the object) is limited at present to the nanoscale, due to quantum and thermal decoherence effects.
Another candidate might be a suspended mirror, \cite{micro}.

Conceivably, the toy set-up (a pointer in an external potential) I used in previous sections for illustrative purposes might be relevant. 
Perhaps it is worth spelling out how testing would go if it could actually be realized. So consider a two-well (quartic) external potential
of macroscopic width $2\,R$, and imagine a cloud of ``particles" initially in a very small band centered at the origin, 
which is also the local maxima (unstable critical point) of the potential. 
Let's regard the cloud as the measuring device
and assume a coupling with some microscopic quantum system 
initially in a superposition that implies forces on the cloud ``particles"
that can send them on a journey to right and/or left. In the quantum regime for the cloud-plus-particle, a cat state will form, 
but the nonlinear energy will prevent it for the classical regime, resulting in a 
normal cloud that is, after some time, either located essentially at position $(-R)$ {\em or} position $(+R)$ where the potential has its minima. 
Let the drop in potential energy between the origin peak and the minima be $\dV$. 
Assuming the energy contributed by the microsystem is small and other internal energies (e.g., attractions between the ``particles" if they are not independent) 
are also small or do not change, then, in the nearly-macroscopic case, for a cat to form 
the nonlinear energy has to be supplied by the external potential. Let $N_c$ be the point at which the cat state has just broken up. Then we can write, approximately, 

\be
N_c\,\dV \approx w\,N^2_c\,R^2,
\ee

\ni which we can take as estimating the free parameter `$w$' by:

\be
w \approx \frac{\dV}{N_c\,R^2}.
\ee

Of course, if no such transition was observed for the range of $N$ we can produce experimentally, then instead of an estimate of $w$ we could only deduce an upper bound.

\section{Discussion} 

The theory proposed here differs from previous wavefunction theories that either incorporate non-linear terms or otherwise aim to eliminate MD. 
For example, in 1966, Bohm and Bub, \cite{bandb}, proposed adding non-linear terms to QM that generate ``basins of attraction" into which the microsystem falls after a measurement. 
This can explain discrete outcomes and wavefunction collapse. However, dissipative dynamics is not Hamiltonian and does not conserve energy.
The theory presented here is Hamiltonian (and does). Moreover, any theory of the Bohm-Bub type may be ruled out by experiments of the type motivated by 
Weinberg's 1989 papers, if it implies non-linear dynamics of small systems. 
(Differences from Weinberg's formulation of NLQM, particularly concerning the wavefunction normalization, are mentioned in the relevant sections.)

In 1986, Ghirardi, Rimini, and Weber proposed a stochastic ``spontaneous collapse" theory that rapidly eliminates MD after it appears, \cite{grw}. 
By contrast, the theory of this paper is purely deterministic: 
nonlinear terms prevent macroscopic superpositions from forming in the first place.
In addition, a special means for eliminating interference between macroscopic pointer positions, or even microscopic ``particle positions" 
in measurement situations (as in a popular theory called ``decoherence"), isn't required.  
That is the job of the apparatus.
For example, in Stern-Gerlach I assumed that the inhomogeneous magnetic field produces two localized wavepackets moving in opposite directions, which in time will render interference impossible.

One possible objection to the energy mechanism for blocking cats is the theoretical possibility of energy cancellations. E.g., if a system made of positive and negative particles can collapse,
that might yield a large negative energy sufficient to cancel against a large positive energy, permitting a cat to form while conserving energy. However, in the linear theory theorems have been proved
about the stability of matter, \cite{lieb}, stating that the system energy is bounded below by a negative constant times $N$, meaning that it could not cancel a positive energy proportional to $N^2$.
Of course, it must be checked that incorporating the nonlinear energy does not abrogate these theorems.

At present, I cannot describe the final state of microsystem-plus-apparatus in detail. 
Solving high-dimensional, non-linear wave equations analytically appears infeasible; nor, because of the
exponential growth of dimensionality of a quantum system with degrees-of-freedom (`$Q$ qubits' requires a Hilbert space of dimension $2^Q$), 
is it possible to simulate the theory on a digital computer.
Thus I have relied for the claims made about measurement on two facts: (a) energy is conserved, making cat-formation impossible; 
and (b) the non-linear terms do not alter the equation of motion of the COM of the apparatus pointer. 
Conceivably, this theory is ``chaotic"; i.e., possesses unstable dependence on initial conditions or measurement parameters, as is common for high-dimensional, nonlinear systems. 

Thus, the question of what determines the final state, as well as the ultimate cause of random outcomes, remains open. 
These topics are addressed in the second paper in this series.

\section{Mathematical Appendix\label{math_app}}

\def\hsotm{\left[\frac{\hbar^2}{2\,m}\right]}
\def\hotm{\left[\frac{\hbar}{2\,m}\right]}
\def\hom{\left[\frac{\hbar}{m}\right]}
\def\Vint{V_{\hbox{int.}}}
\def\ooN{\left[\frac{1}{N}\right]}
\def\trij{\triangle_j}
\def\trik{\triangle_k}
\def\triy{\triangle_y}
\def\ooh{\left[\frac{1}{\hbar}\right]}

Here is the derivation of equation (\ref{ev_eqn}). We begin with an apparatus N-degree-of-freedom (``N particle") plus microscopic-degree-of-freedom \Schseqn, 
pretending as usual that space has one dimension to avoid
cluttered expressions, and omitting the nonlinear term at present:

\bar
\no i\,\hbar\,{\pa\,\psi\,/\,\pa\,t} &\=& \- \hsotm\,\sjN\, \triangle_j\,\psi \- \hsotm\,\triangle_y\,\psi\\
\no  && \+ V_a(x_1,x_2,...)\,\psi \+ \Vint(x_1,...,x_N;y)\,\psi.\\
&& \label{mult_eqn}
\ear

\ni where $\triangle_j = \pa^2/\pa x_j^2$; $\triangle_y = \pa^2/\pa y^2$; $x_j$ stands for an apparatus coordinate; $y$ stands for some microscopic system coordinate; 
$V_a$ stands for apparatus potential energies,
possibly including an external field and internal forces; and $\Vint$ is the interaction potential of macroscopic and microscopic systems. I'll 
discuss forms for the potentials later. 

The problem: study

\be
\ddot{X} \= d^2/dt^2\, <\psi|\,\ooN\,\skN\,x_k\,|\psi>.
\ee

We have

\bar
\no d/dt\, <\psi|\,x_k\,|\psi> &\=& \hotm\{<i\,\sjN\,\trij\,\psi + i\triy\,\psi|x_k|\psi> \\
\no &&  \+ <\psi|x_k|i\,\sjN\,\trij\,\psi + i\triy\,\psi>\} \\
\no &&  \+ \ooh\{\,<  - i\,\left(\,V_a + \Vint\,\right)\psi|x_k|\psi> \\
\no &&  \+ <\psi|x_k|- i\,\left(\,V_a + \Vint\,\right)\psi> \}\\
&&\label{big_eqn}
\ear

In (\ref{big_eqn}), the terms involving the $V$'s cancel, while the terms containing $\trij$ or $\triy$ and $x_k$ with $k \neq j$ can be integrated by parts and disappear, leaving:

\be
 d/dt\, <\psi|\,x_k\,|\psi> \= \hotm\,\{-i\,<\,\trik\,\psi|x_k|\psi> \+ i\,<\psi|x_k|\trik\,\psi>\}.
\ee

Performing two more IBP's yields (as momentum is represented in QM by the operator: $i\,\hbar\,\pa\,/\pa\,x$, this equation just says that the velocity is equal to the momentum
divided by the mass):

\def\ppxk{\frac{\pa}{\pa x_k}}
\def\pppxk{\frac{\pa \psi}{\pa x_k}}
\def\pVapxk{\frac{\pa V_a}{\pa x_k}}
\def\pVintpxk{\frac{\pa \Vint}{\pa x_k}}

\be
 d/dt\, <\psi|\,x_k\,|\psi> \= i\,\hom\,<\,\ppxk\,\psi|\psi>.\label{dt_eqn}
\ee

\def\iom{\left[\frac{i}{m}\right]}

Next:

\bar
\no d^2/dt^2\, <\psi|\,x_k\,|\psi> &\=& \iom\,\{\,<\ppxk\left(\, i\,\hsotm\,\left[\sjN\,\trij\,\psi + \,\triy\,\psi \right] \-\right.\\
\no && \left. i\,\left[\,V_a + \Vint\,\right]\,\psi\,\right)\,|\psi>\\
\no && \+ \,<\pppxk\,| i\,\hsotm\,\left[\,\sjN\,\trij\,\psi + \triy\,\psi\,\right] \-\\
\no && i\,\left[\,V_a + \Vint\,\right]\,\psi>\,\}\\
&&
\ear

IBPs twice on the terms containing $\trij$ and $\triy$, they vanish, as do the terms containing derivatives of $\psi$ times $V$'s; that leaves only the terms with $V'$'s. 
Averaging over $k$ we attain, finally:

\be
 m\,\ddot{X} \= \- \ooN\,\skN\, <\psi|\,\left[\,\pVapxk + \pVintpxk\,\right]\,|\psi>.\label{final_eqn}
\ee

Just to see what this looks like in a special case, we might take:

\def\sjnkN{\sum_{j \neq k}^N}

\bar
\no V_a &\=& \sjN\,V(x_j) \+ \sjnkN\,u(x_j - x_k);\\
\Vint &\=& \alpha\,\left\{\,\sjN\,x_j\,\right\}\,y,
\ear

\ni where $u(\cdot)$ stands for internal energy holding the apparatus needle together and $\alpha$ is an interaction constant. 
It is easily seen that the term involving $u$ disappears from (\ref{final_eqn}) and we get:

\be
\no m\,\ddot{X} \= \- \ooN\,\skN\, <\psi|\,V'(x_k)\,|\psi> - \alpha\,<\psi|\,y\,|\psi>.\label{final2_eqn}
\ee

What about the nonlinear term? It adds a term to the right side of \Schseqn\ of form:

\def\pcHnlppb{\frac{\pa \cHnl}{\pa \psi^*}}
\def\Vnl{V_{\hbox{NL}}}

\bar
\no \pcHnlppb &\=& \left\{\,w\,\left(\,\sjN\,x_j\,\right)^2 \- 2\,w\,<\psi|\sjN\,x_j\,|\psi>\,\left(\,\sjN\,x_j\,\right)\,\right\}\,\psi;\\
&\=& \Vnl(x_1,...;\psi)\,\psi.\label{Vnl_eqn}
\ear

Moreover, $\Vnl$ can be treated (despite containing a dependence on $\psi$) exactly as the other potentials in the derivation of (\ref{final_eqn}), and therefore contributes a term:

\be
2\,w\,<\psi|\left\{ \,\sjN\,x_j - <\psi|\,\sjN\,x_j\,|\psi>\,\right\}\,|\psi>,
\ee

\ni which vanishes.
\vskip0.2in

Concerning the NLQM \Sch's equation, (\ref{geneq}), I define the right side as if the chain rule holds; i.e.,
if $\psi = Q + i\,P$, so that $Q = (1/2)\,(\psi + \psi^*)$ and $P = (-i/2)\,(\psi - \psi*)$, then: 

\bar
\no \frac{\partial\,H}{\partial\,\psi^*} &\=& \frac{\partial \,H}{\partial\,Q}\,\frac{\partial\,Q}{\partial\,\psi^*} \+ \frac{\partial \,H}{\partial\,P}\,\frac{\partial\,P}{\partial\,\psi^*}\\
 &\=& \frac{1}{2}\,\frac{\partial \,H}{\partial\,Q} \+ \frac{i}{2}\,\frac{\partial \,H}{\partial\,P},
\ear

\ni from which (\ref{geneq}) can be written as the pair:

\bar
\ni \frac{\partial\,Q}{\partial t} &\=& \frac{1}{2\,\hbar}\,\frac{\partial \,H}{\partial\,P};\\
 \frac{\partial\,P}{\partial t} &\=& - \frac{1}{2\,\hbar}\,\frac{\partial \,H}{\partial\,Q};
\ear

\ni which, except for the factor of $1/2\,\hbar$, has the form of Hamilton's equations. 
(Of course, $Q$ and $P$ are fake ``generalized coordinates" without units, and NOT the physical position and momentum.) 
In particular, \Sch's original equation can be described mathematically as being 
the special case of an infinite-dimensional, linear Hamiltonian system that commutes (as flows) with the simplest case with 
Hamiltonian $\sum\,[Q_j^2 + P_j^2]$, which generates the overall phase-rotation leaving the quantum state invariant. 
Whether you want to insist on that last part in a nonlinear extension depends on your goals and philosophy.

\vskip0.2in
For the one-dimensional example of ``more-realistic correlations" cited in section \ref{spec_sect}, I assumed a two-level function:

\be
f(x) \= 
\begin{cases}
f_1, \, \hbox{for}\,-\epsilon \leq x \leq +\epsilon;\\
- f_2, \, \hbox{for}\,\epsilon \leq x \leq \eta\,\hbox{or} - \eta \leq x \leq -\epsilon;
\end{cases}
\ee

\ni where $f_1,\,f_2,\,\eta,\,\epsilon$ are positive constants with $f_2 < 1$ and $\eta > \epsilon$. The reason that there are both ``attractive" and ``repulsive" regions 
(meaning where the conditional density increases or decreases) is because
necessarily

\be
\int\int\,dx_1\,dx_2\,\rho_1(x_1)\,\rho_1(x_2)\,f(x_1 - x_2) \= 0.
\ee

Using flat densities, this necessitated by explicit calculation:

\be
\frac{f_1}{f_2} \= \frac{\eta - \epsilon}{\epsilon}.
\ee

Another calculation yields, plugging in the above:

\bar
\no \int\int\,dx_1\,dx_2\,\rho_1(x_1)\,\rho_1(x_2)\,f(x_1 - x_2)\,x_1\,x_2 &\=& \left[\,\frac{L\,f_2\,\epsilon}{6}\,\right]\,\left\{\,\frac{-f_1}{f_2} + \frac{2\,(\eta - \epsilon)}{\epsilon}\,\right\}\\
&\=& \frac{L}{6}\,f_1\,\epsilon.
\ear

\ni which was cited in the text.

\end{document}